\UseRawInputEncoding

\documentclass[12pt]{article}

\usepackage{amsmath, amsfonts, amssymb}
\usepackage{mathrsfs}
\usepackage{theorem}
\usepackage{bm}

\usepackage{circuitikz}
\usepackage{tikz}
\usepackage{ifluatex}
\usetikzlibrary{shapes.geometric, arrows}
\usepackage{pgfplots}

\usepackage{color}
\usepackage{tikz}
\usepackage{pst-node}

{\theorembodyfont{\rmfamily} }
{\theorembodyfont{\rmfamily} }

\def\E{\mathbb E}

\newcommand{\fin}{\hspace*{\fill}\rule{0.3em}{1ex}}

\numberwithin{equation}{section}

\allowdisplaybreaks

\begin{document}

\title{Time-Variant System Reliability  with Infinite Delay Based on Girsanov's Transformation
}

\author{Hussein K. Asker${}$ \footnote{ Email:husseink.askar@uokufa.edu.iq}   \\[0.2cm]
{\small  Department of Mathematics, Faculty of Computer Science and Mathematics, }\\
{\small Kufa University, Al-Najaf, IQ }\\
}
\maketitle

\begin{abstract}
This work addresses the reliability of time variant system appreciation models of dynamic systems, where regulatory equations are expressed as an infinite delay collection of stochastic functional differential equations (SFDEwID). Reliability estimation forms of series and parallel systems tackled depending on  Monte Carlo simulations based on extends Girsanov's transformation for infinite delay SFDEs.

\end{abstract}

\section{Introduction}  

System reliability defines as the possibility that a system (for one component or more) will execute properly in compliance with a specified set of operating conditions for a particular time-interval, see \cite{Barlow}. The reliability of a system changes depending on time. In other words, reliability is a time-variant value. Within the practice of engineering, when the structure's material properties deteriorate in time, time-variant reliability issues appear, like corrosion in steel structures, concrete shrinkage and creep phenomena. Also, if random loading was a random process, such as temperature, wave height, traffic loads \cite{asl}. Estimation the reliability consists of determining probability of achievement  and probability of disappointment in a duration of time. Also, it depends on components configuration where system success (failure)  described as combinations or intersections of the  failure events of these components \cite{ kanjilal2019, sundar}. There are many kinds of component's settings available, such as series, parallel and many others, for details see \cite{kumar}. Any dynamic system represented as a system of equations, but not all system of equations have specific solutions, especially, if the governing equations are stochastic differential equations (SDEs). The Girsanov's transformations applied to such governing system of equations to achieve estimators for system reliability or failure probability, see \cite{kanjilal2019, kanjilal2017, sundar, sudret}. In this chapter, we extended and applied the Girsanov's transformations to a system of infinite delay stochastic functional differential equations (SFDEwID) to get models of estimation of series and parallel system's reliability.

\section{Theoretical background of system reliability}
This section outlines the basic theory of system reliability, and the exposition follows mostly Barlow \cite{Barlow}.
\subsection{Reliability function}

Role of reliability related to the survival of a system in the specified interval of time   $ (0, t] $  which is the chance that the system does not malfunction in the $(0, t)$ period, where $ t $ is the moment when the system is already running and mathematically described as follows:
\begin{equation}\label{r}
P_S(t) = 1 - P_F(t) = 1 - \int_{0}^{t}f(s)ds = \int_{t}^{\infty}f(s)ds
\end{equation}
where, $ P _{F}(t)$ is the risk of failure during the interval (0, t] and the map $ f(s) $ known as the density function of the probability (pdf), thus from \eqref{r} we have $ 0  \le P_S(t) \le 1 $.

\subsection{Failure rate function}

The function of the failure risk (hazard rate function) is the likelihood that a component (system) fails within a given period of time $(t, t+h]$ by recognizing that the component (system) is running at time $t$. This possibility is described as $t$:
\begin{equation}\label{pr}
Pr( t<T \le t +h|T > t) = \dfrac{Pr(t < T \le t +h)}{Pr(T > t)}= \dfrac{P_F(t +h) - P_F(t)}{P_S(t)},
\end{equation}
where $ T $ is failure time. Also, one can obtain failure rate function $ \lambda(t) $ by: 
\begin{equation}
\lambda(t) = \lim\limits_{h\rightarrow 0} \dfrac{Pr(t < T \le t+ h |T >t  )}{h}  = \lim\limits_{h\rightarrow 0} \dfrac{P_F(t +h) - P_F(t)}{h} \dfrac{1}{P_S(t) } = \dfrac{f(t)}{P_S(t)}.
\end{equation}
\subsection{Modelling Failure Rate}
In this work, the normal distribution used to   describe the failure rate. A normal distribution's probability density function (pdf) with failures times of $t$ is set accordingly:
\begin{equation}
f(t) = \dfrac{1}{\sigma \sqrt{2\pi}}e^{\frac{-(t-\mu)^{2}}{2 \sigma^{2}}} 
\end{equation}
where $ \mu $ is the mean value,
$ \sigma $ is the standard deviation and $ \sigma^{2} $  is the variance of the distribution.
The normal distribution's cumulative function (cdf) is expressed as:
\begin{equation}
P_{F}(t) = \phi(\dfrac{t-\mu}{\sigma})= \int_{- \infty}^{t}\dfrac{1}{\sigma \sqrt{2\pi}}e^{\frac{-(s-\mu)^{2}}{2 \sigma^{2}}}  ds,
\end{equation}
thus the reliability is:
\begin{equation}
P_S(t) =1- \phi(\dfrac{t-\mu}{\sigma}).
\end{equation}
Consequently, the failure rate is:
\begin{equation}
\lambda (t)= \dfrac{f(t)}{P_S(t)} = \dfrac{1}{\sigma \sqrt{2\pi}\big[ 1- \phi(\dfrac{t-\mu}{\sigma})\big]}e^{\frac{-(t-\mu)^{2}}{2 \sigma^{2}}}. 
\end{equation}

\subsection{System structure function}
Any system is a set of elements (subsystems); thus, the system structure function depends on the configuration of the elements (subsystems). For a system formed by $ n $ elements, let a vector $ \underline{x} = (x_{1}, x_{2}, \cdots, x_{n})  \in \{0, 1\}^{n} $ with $ x_{i} = 1 $ if the $ i $th element  (subsystem) is in operating state and $ x_{i} = 0 $ if not, be the state vector which provides the status of each element  (subsystem) in the system and $ \Phi = \Phi(x) : \{0; 1\}^{n} \rightarrow \{0; 1\}  $ specify a function of the system structure, where $ \Phi = 1 $ if the function of the system for its corresponding elements  (subsystems) is state vector $ \underline{x} $ and  $ \Phi = 0 $ if not. Now, we are introducing two kinds of structures as an example to explain the structural link between the Structure of the system feature and its elements state vector, for more details see \cite{feng}.

\textbf{1- Series System:} A system that operates if and only if all the elements  (subsystems) operate is referred to as a series system. See Figure 2.4.1. 
\begin{center}
		\setlength{\unitlength}{.5mm}
	\begin{circuitikz}
		\tikzset{quad/.style={draw, thick, minimum height=.8cm, minimum width=.8cm}}
		\node[quad] (A) at (4,0) {1}; 
		\node[quad] (A) at (6,0) {2};
		\node[quad] (A) at (8,0) {$\dots$};
		\node[quad] (A) at (10,0) {$ n $};
		\put(58,0){\line(1,0){14}} 
		\put(208,0){\line(1,0){14}}
		\put(168,0){\line(1,0){24}}
		\put(128,0){\line(1,0){24}}
		\put(88,0){\line(1,0){24}}
	\end{circuitikz}
\end{center}
\begin{center}
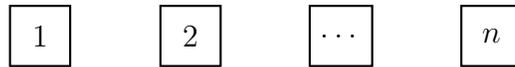
	Figure 2.4.1: System of series with $ n $ elements  (subsystems).
\end{center}
And this system has the structure function
\begin{equation}\label{s1}
 \Phi(\underline{x}) = x_{1} \cdot x_{2} \cdots x_{n} = \prod_{i=1}^n x_{i}. 
\end{equation}

\textbf{2- Parallel System:} A system that runs while at least one element  (subsystem) functions is called a parallel system. The block diagram for this system is seen in Figure 2.4.2.
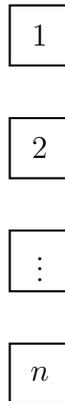
\begin{center}
	\setlength{\unitlength}{.5mm}
	\begin{circuitikz}
		\centering
		\tikzset{quad/.style={draw, thick, minimum height=.8cm, minimum width=.8cm}}
		\node[quad] (A) at (4,0) {$ n $}; 
		\node[quad] (A) at (4,1.5) {$ \vdots $};
		\node[quad] (A) at (4,3) {2};
		\node[quad] (A) at (4,4.5) {1};
		\put(60,0){\line(0,1){90}}  
		\put(100,0){\line(0,1){90}}
		\put(46,50){\line(1,0){14}} 
				\put(100,50){\line(1,0){14}} 
		\put(60,0){\line(1,0){12}}
		\put(60,30){\line(1,0){12}}
		\put(60,60){\line(1,0){12}}
		\put(60,90){\line(1,0){12}}
		\put(88,0){\line(1,0){12}}
		\put(88,30){\line(1,0){12}}
		\put(88,60){\line(1,0){12}}
		\put(88,90){\line(1,0){12}}
			\end{circuitikz}
\end{center}
\begin{center}
	Figure 2.4.2: Parallel structure with $ n $ of elements.
\end{center}

The function of the system structure is provided by
\begin{equation}\label{s2}
 \Phi(\underline{x}) = 1 - (1 - x_{1})(1- x_{2}) \cdots (1- x_{n}) = 1 - \prod_{i=1}^n (1-x_{i}). 
\end{equation}


\subsection{System reliability function}

The relationship within system's components is essential to evaluate  the system reliability as a whole, and thus once the function $\Phi(\underline{x})$ of the system structure is identified then  the reliability can be determined. For independently functional components,  if $ p_{i} $ is the reliability of element $ i $, and $ R=P_S $ is the reliability (probability of success) of the system then the reliability of the series system, depending to \eqref{s1}, is:
\begin{equation}
\begin{split}
P_S = P_{S}(\Phi(\underline{x}) = 1)& = P(\prod_{i=1}^n x_{i}= 1) = P(x_{1} = 1, x_{2} = 1, \cdots , x_{n} = 1)\\ & =\prod_{i=1}^n  P(x_{i} = 1) = \prod_{i=1}^n p_{i} 
\end{split}
\end{equation}
 and, similarly, a parallel system's reliability where it's structure function shown in \eqref{s2}, is given by
\begin{equation}
\begin{split}
P_{S}
& =1- \prod_{i=1}^n (1-p_{i}) 
\end{split}
\end{equation}


In order to get an estimation of system reliability measure, we suppose that a dynamical system represented as an SFDEwID, then the probability measure will be changed to another absolutely continuous probability measure, and this called the Girsanov's transformation.Where, by adding a control function (drift function) to the noise expression, the adjustment is achieved \cite{okse}, as we explain in the next section. 

\section{SFDEwID and Girsanov's transformation }
 In order to estimate system reliability, Kanjilal \cite{kanjilal2019, kanjilal2017} and Sundar \cite{sundar} have considered a category of dynamical systems controlled by SDEs. In this section, we extended the governed equation to SFDEwID and applied the Girsanov's transformations to get models of estimation of the system's reliability ( in series and parallel modes). Consider a system of SFDEwID
\begin{equation}\label{sys}
 \left\{ \begin{array}{rcl}
d[x(t) ] = b( x_{t})dt + \sigma (x_{t})dw(t) \quad  \mbox{for} \quad t \in [0, T] , &\\
x(0)= x_{0} = \xi = \xi(\theta) :-\infty<\theta\leq0  \in C_{r}  .
\end{array}\right.
\end{equation}

Here $ x(t) $ is a $ n \times 1 $ vector and $ x_{t} =  x(t+\theta) : - \infty < \theta \leq 0 $. $ b( \cdot) $ is a non-negative $ n \times 1 $ drift vector, $ \sigma (\cdot) $ is a $ n \times m $ matrix of diffusion coefficients and $ w(t) $  is an $ m $-dimensional Brownian motion. It is clear that the system \eqref{sys} is a NSFDEwID with the neutral term $ D(\cdot)=0 $. Let $(\Omega,\mathcal{F}, \mathcal{P}) $  be a complete probability space with a filtration $ {\{{\mathcal{F}_{t}\}_{t\in [0,+\infty)}}} $ satisfying the usual conditions. 
 The basic assumption is that performance and design requirements of a dynamical system restrict the significance level of the safe domain response and the dimension parameter $ n $ depends on the problem under study \cite{olsen}.

Let $g(x(t)) $ be a scalar measure (function of limit state ) of a system, and $g^{*} $ be its safe limit value   on $g(x(t)) $. Thus, it is possible to write the failure event as: 
\begin{equation}\label{f}
 F =\{\omega: g^{*}- \max_{- \infty \le t \le T}g ( x(t,\omega)) \le 0\} ,
\end{equation}
 such that, the time interval of interest is $ -\infty < t \le T $. Assume 
 \begin{equation*}
  P_{S}(T) = P\{\omega:\max_{- \infty \le t \le T}g ( x(t,\omega)) \le g^{*}\}  
 \end{equation*} for all $ -\infty < t \le T $. $ P \{\cdot\}$ is a probability measure (reliability measure), that the  performance of the system shall remain
below the safe limit $g^{*} $ for whole times during the given period $ t \in ( - \infty, T]  $.
Therefore, $ P_{S} $ stands for  reliability and
 \begin{equation}\label{mt}
  P_{F}(T) = 1 - P_{S}(T)=  P\{\omega:\max_{- \infty \le t \le T}g ( x(t,\omega))>g^{*}\}, 
\end{equation} 
the probability of failure. In other words the events  \big($ g^{*}-g_{m}  < 0  $\big) and    \big($ g^{*}- g_{m}  > 0 $\big)  represent, respectively,  failure and reliability of the system, where $ g_m = \max_{- \infty \le t \le m}g(x(t))$.  A direct estimator  for reliability of the system $ ( P_{S}(T)) $ by  Monte Carlo method,
in terms of random draws $  x^{j}(t), j=1, \cdots, N $ of $ x(t) $  that we can get by solving   \eqref{mt} numerically, Is provided by

\begin{equation}\label{rff}
\widehat{P}_S(T)=1-\widehat{P}_F(T) =1- \dfrac{1}{N} \sum_{j=1}^{N}  I_{ \Big\{   g^* - \max_{- \infty < t \le T}g(x^{j}(t))  \le 0\Big\}}.    
\end{equation}
It can be shown that   
\begin{equation}
\begin{split}
\E_{\mathcal{P}}[ \widehat{P}_F(T)]& = \E_{\mathcal{P}} \Big[ \dfrac{1}{N} \sum_{j=1}^{N}  I_{ \Big\{   g^* - \max_{- \infty < t \le T}g(x^{j}(t))  \le 0\Big\}}\Big]\\
& = \dfrac{1}{N} \sum_{j=1}^{N}  \Big[P \Big\{   g^* - \max_{- \infty < t \le T}g(x^{j}(t))  \le 0\Big\}\Big]= P_F(T)
\end{split}
\end{equation}
where  $ \E_{\mathcal{P}}[ \cdot]  $ is the expectation under the measure $ \mathcal{P} $ and

\begin{equation}
\begin{split}
	Var(\widehat{P}_F(T)) &=\E_{\mathcal{P}}[\widehat{P}_F(T)]^{2}- \Big(\E_{\mathcal{P}}[\widehat{P}_F(T)]\Big)^{2}\\
	&=\E_{\mathcal{P}}[\widehat{P}_F(T)]^{2}- P_F^{2}(T).\\	  
\end{split}
\end{equation}
Due to 
\begin{equation}
\begin{split}
 \E_{\mathcal{P}}[\widehat{P}_F(T)]^{2}& = \E_{\mathcal{P}} \Big[ \dfrac{1}{N} \sum_{j=1}^{N}  I_{ \Big\{   g^* - \max_{- \infty < t \le T}g(x^{j}(t))  \le 0\Big\}}\Big]\\
 & = \dfrac{1}{N^{2}} \sum_{j,i=1}^{N} \E_{\mathcal{P}} \Big[I_{ \Big\{   g^* - \max_{- \infty < t \le T}g(x^{j}(t))  \le 0\Big\}} \times I_{ \Big\{   g^* - \max_{- \infty < t \le T}g(x^{i}(t))  \le 0\Big\}\Big]}\\   
 & = \dfrac{1}{N^{2}}\Big[ \sum_{j=i=1}^{N} P  \Big(\Big\{   g^* - \max_{- \infty < t \le T}g(x^{j}(t))  \le 0\Big\}\Big) \\
 &\qquad + \sum_{j\ne i=1}^{N}P \Big(\Big\{   g^* - \max_{- \infty < t \le T}g(x^{j}(t))  \le 0\Big\}\Big)P\Big(\Big\{   g^* - \max_{- \infty < t \le T}g(x^{i}(t))  \le 0\Big\}\Big)\Big]\\ 
 & = \dfrac{1}{N^{2}}\Big[ NP_F(T) + \sum_{j\ne i=1}^{N}P_F^{2}(T) \Big] = \dfrac{1}{N^{2}}\Big[ NP_F(T) + (N^{2}-N)P_F^{2}(T) \Big],\\	  
\end{split}
\end{equation}
thus
\begin{equation}
\begin{split}
Var(\widehat{P}_F(T)) &= \dfrac{1}{N^{2}}\Big[ NP_F(T) + (N^{2}-N)P_F^{2}(T) \Big]-P_F^{2}(T) \\
&=\dfrac{P_F(T)(1-P_F(T))}{N}\\	  
\end{split}
\end{equation}
and
\begin{equation}
\lim\limits_{N \rightarrow \infty }Var(\widehat{P}_F) = \lim\limits_{N \rightarrow \infty }\dfrac{P_F(1-P_F)}{N} \rightarrow 0
\end{equation}
 which means, to obtain acceptable estimates of $ P_F $, sample size needed would be enormous \cite{  kanjilal2019, nayek, sundar}. So, to manage the sample size, the recourse is the method of Girsanov’s transformation. By reconstruct the drift term in the system \eqref{sys} via an additional control force $ u(\widetilde{x}_{t}) $ leads to obtain a modified dynamical  SFDEwID system governed by
\begin{equation}\label{sys2}
\left\{ \begin{array}{rcl}
d[\widetilde{x}(t) ] = b( \widetilde{x}_{t})dt + \sigma (\widetilde{x}_{t})\big( u(\widetilde{x}_t)dt + d\widetilde{w}(t) \big) \quad  \mbox{for} \quad t \in [0, T], &\\
\widetilde{x}(0)= x_{0} = \xi = \xi(\theta) :-\infty<\theta\leq0  \in C_{r} ,\qquad
\end{array}\right.
\end{equation}
where $ u(\widetilde{x}_t) $ is an additional drift term of dimension $ n \times 1 $, and $ \widetilde{w}(t) $ is an It\^o's process (stochastic integral) donated by
\begin{equation}
  d\widetilde{w}(t)=-u(\widetilde{x}_t)dt+dw(t); \quad \widetilde{w}(0)=0; \quad t \ge 0.
\end{equation}
The conversion of equation \eqref{sys} to \eqref{sys2}  basically alters  measure $ \mathcal{P} $ to a new probability measure $ \mathcal{Q} $, in this way, $ \mathcal{Q} $ is absolutely continuous  with regard to $ \mathcal{P}$ $ (\mathcal{Q} \ll \mathcal{P}) $. $ \widetilde{w}(t)  $ is a $ m $-dimensional Brownian process with the new probability measure $ \mathcal{Q} $ defined on $ (\Omega, \mathcal{F}) $. By the virtue of Girsanov’s theorem,  the related Radon-Nikodym's derivative that we can explicitly compute , see  \cite{okse},  is provided by
\begin{equation}\label{radon}
\dfrac{y (t)}{y_0} = \dfrac{d\mathcal{P}}{d\mathcal{Q}}(t) = exp \Bigg(-  \int_{0}^{t} u_j(\widetilde{x}_s) d\widetilde{w}_j(s)  - \dfrac{1}{2}  \int_{0}^{t} \big(u_j(\widetilde{x}_s) \big)^{2} ds \Bigg)
\end{equation}
Hence, one now has
\begin{equation}\label{pf}
\begin{split}
\widetilde{P}_F = \int_{F}d\mathcal{P}&= \int_{F}\Bigg( \dfrac{d\mathcal{P}}{d\mathcal{Q}}(T)\Bigg)d\mathcal{Q}\\
&= \int_{\Omega}\Bigg( \dfrac{d\mathcal{P}}{d\mathcal{Q}}(T)\Bigg) I _{F}d\mathcal{Q}= \E_\mathcal{Q} \Bigg[ \dfrac{y (t)}{y_0} I_{\Big\{   g^* - \max_{- \infty \le t \le T}g(\widetilde{x}(t))  \le 0\Big\}}     \Bigg],
\end{split}
\end{equation}
where $ I_{\{\cdot\}} $ is a characteristic function. The equation \eqref{radon}  shows that $ \mathcal{Q} \big[\dfrac{y(t) }{y_0}  \ge 0 \big]=1 $      and   $ \E_{\mathcal{Q}} \big[\dfrac{y(t) }{y_0}  \big]=  \int   \dfrac{d\mathcal{P}}{d\mathcal{Q}}(t) d\mathcal{Q}=1 $ . Through rewriting equation \eqref{radon} as $ y(t)=y_0  f (G(t))$ with $ f (G(t))= exp \Big( G(t)  - \dfrac{1}{2} \int_{0}^{t} \big(  u (\widetilde{x}_s)\big)^2 ds \Big) $ and $ G (t) = - \int_{0}^{t} u(\widetilde{x}_s) d\widetilde{w}(s) $, now, by using the It\^o rule to differentiate, it can be seen that

\begin{equation}\label{dy}
dy(t) = - y(t)  u(\widetilde{x}_{t})d\widetilde{x}(t); \qquad y(0)=y_0.
\end{equation}
 Dependent on equation \eqref{pf}, an estimator for $ P_{F} $ can now be obtained as

\begin{equation}\label{pff}
\overline{P}_F = \dfrac{1}{N} \sum_{j=1}^{N} \dfrac{y^j(T)}{y_0^j} I _{\Big\{   g^* - \max_{- \infty < t \le T}h(\widetilde{x}^{j}(t))  \le 0\Big\}}    
\end{equation}
where $ \widetilde{x}^{j}(t) $ and $ y^{j}(t) $  are obtained as sample solutions of equations \eqref{sys2} and \eqref{dy} respectively with $ y^{j}_{0}\neq0 $ for all $ j =1, ..., N $.\\

\section{Estimation models of system's time-variant reliability  based on component configurations}
 For highlight the aim of determining the control of Girsanov  and the corresponding derivative of Radon-Nikodym  in the time-variance analysis problem, it is useful to introduce examples of estimation's models of systems (in series and parallel mode) reliability. For more configurations see \cite{kanjilal2019, sundar}.
 
\subsection{Reliability of series system}\label{s41}
If $ N $ number of failure components arranged in series then the system failure events, according of \eqref{f}, is given by:

\begin{equation}\label{fi}
F_{j} =\{g^{*}_{j}- \max_{- \infty \le t \le T}g_{j}(x(t)) \le 0\} ,
\end{equation}
where $ j=1, \cdots, N $ and the Probability of failure will be
\begin{equation}\label{pfi}
P_{F} = \Big( \bigcup_{j=1}^{N} \{g^{*}_{j}- \max_{- \infty \le t \le T}g_{j}(x(t)) \le 0\}\Big) .
\end{equation}
Consequently, by the equations \eqref{rff} and \eqref{pff} the estimators for series system reliability given by:
 
\begin{equation}\label{rffs}
\widehat{P}_S=1- \dfrac{1}{N} \sum_{j=1}^{N}  I_{ \Big( \bigcup_{j=1}^{N} \{g^{*}_{j}- \max_{- \infty \le t \le T}g_{j}(x(t)) \le 0\}\Big)},    
\end{equation} 
 and
 
\begin{equation}\label{pffrr}
\overline{P}_S = 1-\dfrac{1}{N} \sum_{j=1}^{N} \dfrac{y^j(T)}{y_0^j} I_{ \Big( \bigcup_{j=1}^{N} \{g^{*}_{j}- \max_{- \infty \le t \le T}g_{j}(x(t)) \le 0\}\Big)}.    
\end{equation}

\subsection{Reliability of parallel system}

By the same way above, If $ N $ is the total number of failed components arranged in parallel, then the structure  probability failure is specified by
\begin{equation}\label{pfii}
P_{F} = \Big( \bigcap_{j=1}^{N} \{g^{*}_{j}- \max_{- \infty \le t \le T}g_{j}(x(t)) \le 0\}\Big)
\end{equation}
where the estimators for parallel system reliability are:

\begin{equation}\label{rffpp}
\widehat{P}_S=1- \dfrac{1}{N} \sum_{j=1}^{N}  I_{ \Big( \bigcap_{j=1}^{N} \{g^{*}_{j}- \max_{- \infty \le t \le T}g_{j}(x(t)) \le 0\}\Big)},    
\end{equation} 
and

\begin{equation}\label{pffr}
\overline{P}_S = 1-\dfrac{1}{N} \sum_{j=1}^{N} \dfrac{y^j(T)}{y_0^j} I_{ \Big( \bigcap_{j=1}^{N} \{g^{*}_{j}- \max_{- \infty \le t \le T}g_{j}(x(t)) \le 0\}\Big)}.    
\end{equation}

\end{document}